# Detection of Cancer Stages via Fractal Dimension Analysis of Optical Transmission Imaging of Tissue Micro Arrays (TMA)

**Shiva Bhandari[1], Sri Choudannavar[1, 2], Ethan Avery[1], Peeyush Sahay[1] and Prabhakar Pradhan[1*]**

[1.] *BioNanoPhotonics Laboratory, Department of Physics and Materials Science,*
*The University of Memphis, Memphis, TN38152, USA*

[2.] *CRESH Summer Program, The University of Memphis, Memphis, TN38152, USA*

*\*Corresponding author: ppradhan@memphis.edu*

## ABSTRACT

Cancer is an epidemic worldwide. At present one in four persons has cancer and this statistic will change to one in a two person in the near future. It is now known that war against cancer is the early, curable detection and treatment. Affordable, quick and easy detection methods are, therefore, essential. Standard pathologist way of detecting cancer is looking at the stained biopsy tissue samples under microscope, brings lots of human error. A tissue is a spatial heterogeneous medium and it has fractal properties due to its self-similarity in mass distribution. It is now known that with the progress of cancer the tissue heterogeneity changes due to more mass accumulations and rearrangement of intracellular macromolecules such as DNA and lipids etc. Furthermore, there are tissue micro array (TMA) samples available that provides array of hundred samples in one glass slides. In this study, using reflectance microscopy we have analyzed the fractal dimension of 5µm colon TMA samples to correctly distinguish between normal, adjacent to cancer, benign, stage-1 and stage-2 colon cancers. This fractal property of the tissues is also supported by entropy and spatial correlation calculations. The application of this method for diagnostic applications is also discussed.

Keywords: Cancer, TMA, Fractal dimension, Correlation length, Entropy, colon cancer

1. **INTRODUCTION**

Cancer is a disease characterized by an unchecked division, proliferations, and survival of abnormal cancerous cells. Cancer starts due to the several genetic mutations of the DNA and RNA and followed by the proliferation and metastasis [1]. Cancer is an epidemic [2] and to fight its menace, it is importance to have early as well as accurate detection of the stages of cancer [3]. As a result, an affordable, efficient and accurate detection of different stages of cancer is always desired. Most of the diagnosis is still done by visual examination of radiological images, microscopy of stained biopsy specimens, direct observation of tissues under the microscope by pathologist, and so on. The microscopy images are qualitatively analyzed by an expert pathologist who tries of manually identify the abnormal features, e.g., high indices of mitosis, structural irregularities, etc., in the cells and tissues to detect cancer [4]. As these examinations are manually done, they are prone to human error of missing the correct diagnosis. Additionally, such an approach lacks accuracy in identifying the different stages of cancer. Therefore, more quantitative approaches are desired in cancer diagnostics studies [5].

Biological tissues are semi porous dielectric/refractive index media with spatial heterogeneity in their mass density distribution. It has been now known that most tissues have self-similar structure and therefore can be analyzed in terms of fractal dimensions [6]. The progress of cancer results in more mass accumulation inside the cells, rearrangement of the DNA, RNA and heterochromatin, as well as a change in the structure of the extra cellular matrix. Such changes alters the spatial structure of the tissues which may vary from nano- to micron length scales, leading to changes in spatial heterogeneity as well as the fractal nature of the tissues [7]. Consequently, the fractal dimension is considered as a useful parameter to characterize cancer cells and tissues. Fractal dimension analysis essentially serves as a morphometric measure of the irregular structures such as tumor. Therefore, fractal dimension analysis performed for different cancer samples may allow to quantify the degree of tumorigenicity of these samples [8]. In this work, we conducted fractal dimension analysis of normal and different stages of colon cancer tissues using tissue micro arrays (TMA) samples. Tissue micro array is a rapidly evolving diagnostic and research tool for tissue samples organization on glass slides. The slide consists of small tissue samples of numerous different cases assembled on a single histologic slide in a systematic array (see Section 2.B), which allows high throughout analysis of multiple specimens [9, 10]. Importantly, in TMA all the tissue samples are of fixed specifications, for example the one used in the present study has samples of 5 μm thickness and 1.5 mm of diameter. Therefore, another advantage of using TMA is that specifications of any measurement technique would be generic to such micro array parameters. The choice to study colon cancer case was made upon considering its distinct significance as one of the most common cancer found in United States (US). In fact, colon cancer is the third most commonly diagnosed cancer in both men and women in US. One in 22 men and one in 24 women will be diagnosed with colon cancer in their lifetime [11]. We used colon TMA containing different grades of cancer, including normal colon tissue sample to compare the results.

Towards the end, to support our fractal results, we have also performed entropy and correlation analysis of all the samples and have compared the results with the results of fractal dimension analysis.

## 2. METHODS

**A. Mathematical methods:**

*Fractal and Fractal dimension:* A fractal is a structure which exhibits self-similar structural patterns at different length scales and are often characterized in terms of a mathematical parameter called 'fractal dimension' [10,12]. The typical examples of 2D fractal patterns are shown in Fig. 1 (generated using MATLAB software). The fractal system shown in Fig. 1(a) is called carpet deterministic fractal, while the one represented in Fig. 1(b) is known as random fractal pattern generated by using diffusion limited algorithm (DLA).

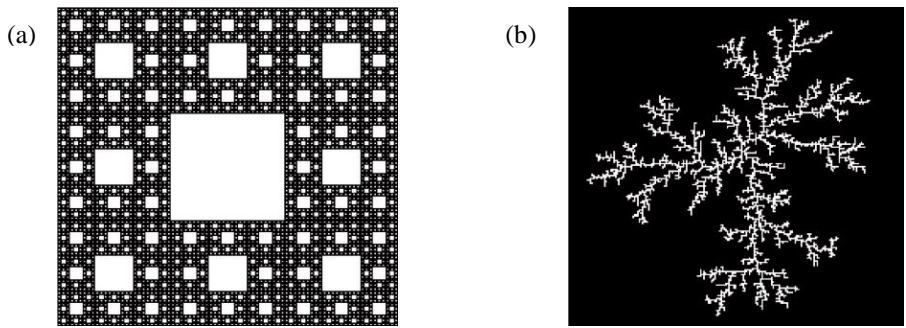

Fig.1: Deterministic and random fractal: (a) The Sierpinski Carpet (deterministic fractal), generated in Matlab with a fractal dimension of 1.8928. (b) Typical example of random fractal generated by the DLA simulation algorithm which has fractal dimension of $D_f$=1.65.

It should be noted that natural biological samples are more akin to random fractal patterns, where the self-similarity at different length scales can be seen upon statistically analyzing their structure. Box counting is a commonly used method to characterize such fractal patterns by measuring their fractal dimensions [13, 14]. In this method, the fractal is imagined to be lying on an evenly spaced grid and the number of boxes required to cover the fractal structure are counted. The fractal dimension is calculated by noting down how this number changes as we make the grid finer by applying a box counting algorithm. The fractal dimension is essentially an index for characterizing fractal patterns by quantifying their complexity as a ratio of change in detail to the change in scale. It can also be considered as a measure of the space-filling capacity of a pattern. If $N(r)$ is the number of the boxes of side length $r$ required to cover the fractal structure, then the fractal dimension is defined as [15]:

$$D_f = \ln(N(r))/\ln(1/r) \tag{1}$$

In case of the random nature of the structure, we take an average of the many realizations of the fractal structure to get the average fractal dimension $D_f$.

The general aim of our research is to investigate the possibility of standardizing numerical index values to characterize different colon tumors, e.g., through fractal dimension analysis, without the use of additional expensive stains, reagents and invasive procedures and so for this purpose colon TMA samples were used.

*Spatial correlation of the mass density fluctuations in tissues:* The spatial correlation of the mass density fluctuations that is corresponding to the intensity fluctuations depends on the varieties of the parameter. As the system is fractal, one expects power law correlation decay or long-range correlation. However due to the finite size effect we will calculate the correlation as a weakly varying decay function and match with an exponential decay form. The function can be written for two refractive index fluctuations points at $r$ and $r'$ as $n(r)$ and $n(r')$ respectively:

$$<dn(r)*dn(r')> = <dn^2> \times exp(-|r-r'|/l_c) \tag{2}$$

Where $l_c$ is the spatial weak correlation decay length scale for the refractive index fluctuations.

*Entropy of the structural disorder*: Given a structure, we can measure the structural disorder via measuring the entropy (*S*) defined as follows:

$$S = -\sum_{r} n(r) \times \log n(r) \qquad (3)$$

Where *n(r)* in Eq. 3 denotes bins for logical arrays in the images. Entropy provides a measure of randomness of the sample, and thus allows to quantify the texture of the sample.

We have used the above three mathematical parameters, namely the fractal dimension, correlation length, and entropy measurement to perform the analysis to characterize the colon samples.

**B. Tissue Samples and Image Collection**

*TMA colon tissue samples*: A colon cancer TMA (CO808, Biomax, USA), shown in Fig. 2 (c), was used as the colon tissue sample to study in this work. Out of 80 total colon cores in the TMA, only 40 were selected for imaging. Selected colon tissue samples ranged from normal and different stages of cancer as follows; 1) normal tissue, 2) benign tumors, 3) stage 1 malignant tumors, 4) stage 2 malignant tumors, and 5) cancer adjacent tissue. Each core in the TMA has a diameter of 1.5 mm and thickness of 5 µm.

*Transmission optical microscopy*: Optical microscopy images were acquired using Olympus BX50 Microscope (Olympus, USA) and an Infinity2 Microscopy Camera (Lumenera Corp., Canada) that is attached to the microscope head to capture the images, as shown in Fig. 2 (a). The TMA samples were kept on the sample platform of the microscope, being operated in transmission mode, to acquire the images. The basic principle how the images are taken from the microscope is shown in diagram in Fig. 2 (b). The images were collected with the Lumenera CAPTURE software.

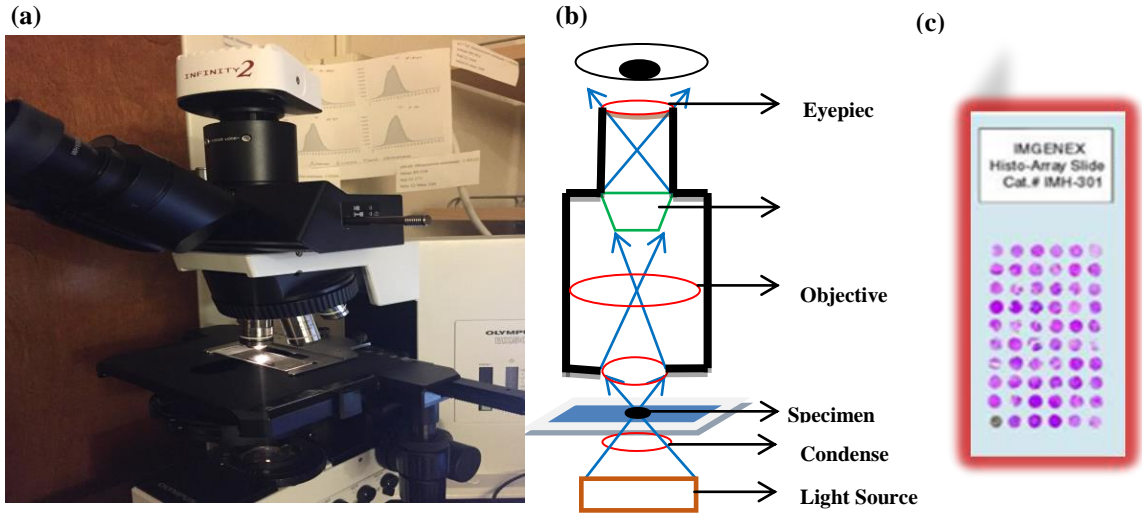

Fig.2: (a) Olympus BX50 Microscope and an Infinity2 Microscopy Camera head. (b) Light transmission Diagram for the microscope. This schematic figure shows the basic principle how light is transmitted in the microscope used. (c) Tissue Micro Array Sample.

Several images (9-12) of each core were taken to achieve an accurate representation of the tissue samples. Images at magnification ×100 were sampled and digitized to eight bits of gray level and stored as digital files.

It is now known that the mass density *dm(r)* at a point inside the tissue and the refractive index *n(r)* at that point are linearly related to each, such that $n(r) \propto dm(r)$ [16]. In case of transmission microscopy, the transmission intensity is proportional to the scattering average refractive index or n(x). Then using the transmission equation, we can write:

$$I(z) = I_0(z) * \exp(-\mu\, n(x)\, z) = I_0(z) * [1 - \mu\, n(x)\, z] \tag{4}$$

Where *I (z)* is the transmission intensity of the micro array along the vertical direction μ is the scattering coefficient of the disordered tissue media. Therefore, the vertical (z-direction) intensity transmission of the voxel at a point *r* provides functional relationship with the refractive index or mass density.

## C. Image Analysis

The collected images were processed by ImageJ software. The fractal dimension of each image was calculated via the box counting method. Each tissue sample received an averaged value for each of the parameters in the 9-12 images taken for each type of cancer cases. The values were then further plotted in Excel for the statistical analyses of these data.

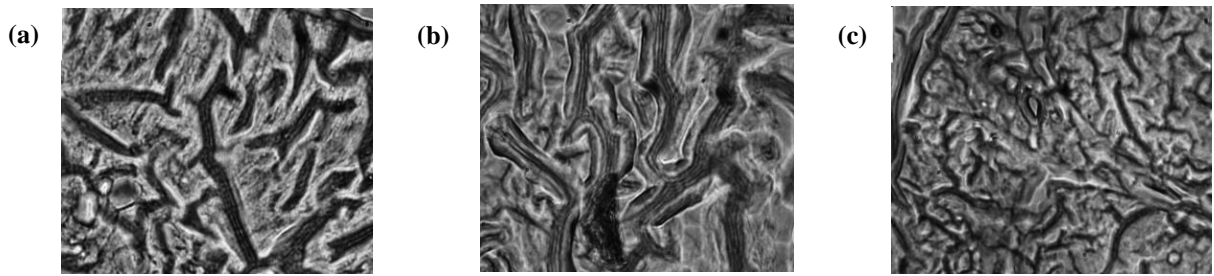

Fig.3: Gray scale microscopic images of different cell types. (a) normal tissue type, (b) stage 1 cancer tissue type, (c) stage 2 cancer tissue type.

## 3. RESULTS

The representative microscopic images of the normal, stage 1 and stage 2 cell types are shown in Fig.3. These images are then further processed by ImageJ (obtained from *https://imagej.nih.gov*) to find the fractal dimension.

Fractal dimension was calculated for five different cases, i.e., normal, adjacent, begin, stage-1, and stage 2 categories. The results are shown in Fig.4. The bar graph result shows that fractal dimension value increases with the increase in the cancer stage. The normal tissue sample had the lowest fractal dimension, followed by the tissue sample adjacent to tumor. Subsequently, begin tumor, stage1, and stage 2 cancer samples had fractal dimensions in increasing order. The actual fractal dimension values recorded are: 1.7192 for normal tissues, 1.7560 for adjacent tissues, 1.7720 for benign tumors, 1.8820 for stage-1 cancer tissues and 1.9336 for stage-2 cancer tissues. Student's t-test for each pair of measurement obtained p-value < 0.05, suggesting that the fractal dimension values for all the different samples are different. This result can be understood in the light of the fact that the progress of cancer results in more mass accumulation in the tissue as well as to the extra cellular matrix [17]. This leads to more filling in the tissue sample, and thus increases the fractal dimension.

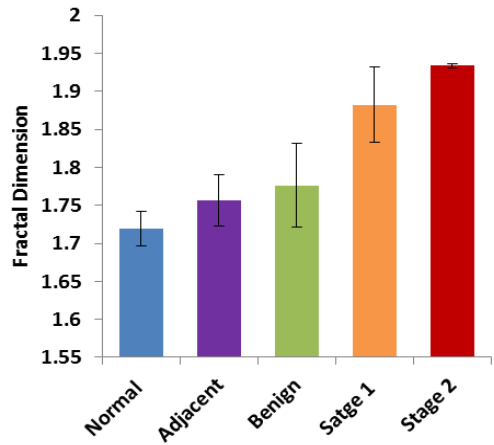

Fig.4: Average Fractal Dimension of different cell types. The fractal dimension is increasing with the increase in the extent of cancer.

To support the results obtained by fractal dimension analysis, we further calculated the spatial correlation length of the mass density fluctuation, and thus refractive index fluctuation of the sample. As pointed out above, the intensity fluctuations in the transmission mode microscopy image can be attributed to the refractive index fluctuations at those points inside the sample. Since, the local refractive index inside a tissue sample is directly proportional to the local mass density inside the tissue, the intensity variation map in the 2D image of the sample can be considered as a representative mass density variation map of the sample. Thus, by analyzing the correlation of the intensity variation in the sample image, one can obtain the spatial correlation of mass density/refractive index variation in the sample. We calculated the spatial correlation decay lengths of the intensity variations of the 2D sample images, with exponential correlation between the two points as pointed out in Section 2. The results are shown in Fig. 5. The result shows that the correlation length increases with the progress of cancer. This result is consistent with the fractal dimension analysis. Increase in correlation length, with the progression of cancer, can be attributed to the filling of the space in the tissues due to the accumulation of mass, as the tumor grows with the progression of the cancer.

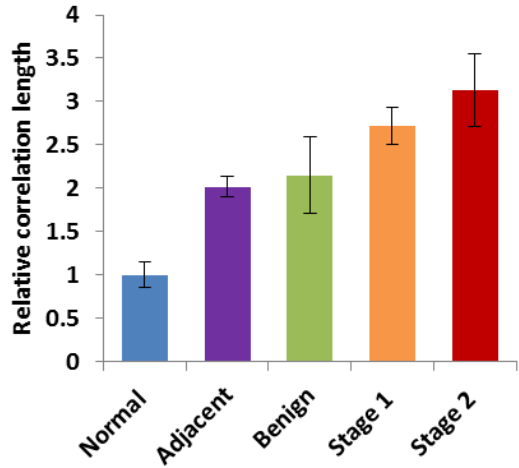

Fig.5 Relative Correlation Length for different tissue types. Correlation length increases with the cancer.

Correlation length provides the nature of the fluctuations and has an effect on disorder, however, we have matched it with the weak exponential decay therefore this is an approximate result. Our results predict that the fluctuations may be higher; therefore, we also calculated the entropy to evaluate the randomness of the sample.

We calculated the entropy from the intensity map of the tissue samples using the entropy method as described in the Section 2. The results are shown in Fig.6.

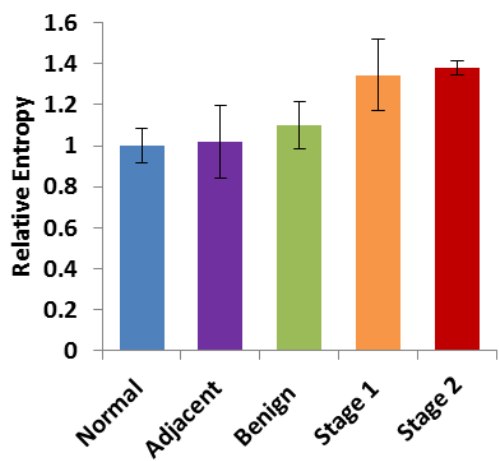

Fig.6: Relative Entropy for different tissue types. Entropy is maximum for the Stage 2 cancer cell type.

Entropy is a measure of randomness and it increases with the progress of cancer. The results in Fig. 6 are consistent with the results of fractal dimension and correlation length. As the randomness increases, there is more fluctuation in mass density due to the fractal nature of the mass density distribution. As the mass density fluctuations increases, the disorder level of the tissue increases and so the entropy. The fractal dimension, correlation length and entropy for the normal and different stages of colon cancer are listed in the Table 1. The increase in the correlation length and entropy with the increase in the tumorigenicity is found consistent with the fractal dimensions.

| Image Feature | Normal Cell Image | Adjacent cell Image | Benign cell Image | Stage1 cell Image | Stage2 cell Image |
|---|---|---|---|---|---|
| Fractal dimension | 1.7192 | 1.7560 | 1.7720 | 1.8820 | 1.9336 |
| Relative Correlation length | 1 | 2.0178 | 2.1495 | 2.7214 | 3.1318 |
| Relative Entropy | 1 | 1.0183 | 1.1010 | 1.3453 | 1.3808 |

Table 1: Different Image Features for different tissue types

## 4. CONCLUSIONS AND DISCUSSIONS

The fractal dimension analysis demonstrated an efficient way to distinguish normal and cancerous tissues of different stages, using a simple transmission intensity using a simpler microscopy. Our results show that the cancerous colon tissues are more fractal in nature compared to the normal and benign tissues. The hierarchy of the fractal dimension is as follows: normal < adjacent to the cancer tissue < benign < stage 1 < stage 2. The gray scale averages obtained from histograms indicates that the degree of disorganization of the tissue reduces the light transmissivity of the tissue, which directly relates to the stage of cancer/aggressiveness of the colon tissue or the potential of the tissue to become cancerous. The normal tissue data shows how deviations in the grayscale values are miniscule, supporting the unorganized and unregulated behavior of the tumor growth. The microscopic images show that stage 2 cancer tissue appears much less organized, filled, and distinct, compared to the normal tissue. We were able to distinguish different stages of colon cancer cases via fractal dimension analysis of colon TMA samples using optical transmission images.

Conventionally, the first hand diagnosis of cancer is done either by qualitative analysis of the microscopic images of the tissue samples. Tissue samples are collected via biopsy and are viewed under microscope by an expert pathologist to spot any structural irregularities in the sample in comparison to a normal healthy sample. Such an approach is highly prone to human error of missing the correct diagnosis. Similarly, in chemical analysis the biopsy samples are stained using specific antibodies and biomarkers and then studied for specific markers. This method is time intensive and is expensive as well. Additionally, both these methods not being quantitative lack accuracy in quantifying the stages of the cancer. In such a scenario, using the fractal analyses, as shown in this work with colon TMA, it may be possible to analyze colon cancer stages with more reliable numerical number and much faster by avoiding the need for special preparations of the samples. Additionally, since the fractal analysis is based on computational calculation, with this method human error occurring in cancer stage detection can also be avoided. Fractal data from cancer research can be complied into a database for doctors to compare patient sample data, and then diagnose cancers more precisely and efficiently. Using TMA in this work further paves a way to perform comprehensive study on a variety of cancer cases with much ease, to gain confidence before doing actual clinical trials and improve results.